\documentclass[journal]{IEEEtran}

\usepackage{amsmath,dsfont}
\usepackage{amssymb,amsthm}
\usepackage{bm}
\usepackage{graphicx}
\usepackage{mathtools}
\usepackage{acronym}
\usepackage{color}
\usepackage{enumerate}
\usepackage{algorithm}
\usepackage{algpseudocode}
\usepackage{algorithmicx}
\usepackage{bm}
\usepackage{soul}
\usepackage{cite}
\usepackage{multirow}
\usepackage{xr}
\makeatletter
\newcommand*{\addFileDependency}[1]{
  \typeout{(#1)}
  \@addtofilelist{#1}
  \IfFileExists{#1}{}{\typeout{No file #1.}}
}
\makeatother

\newtheorem{theorem}{Theorem}[section]

\acrodef{cfar}[CFAR]{constant false-alarm rate}
\acrodef{cpi}[CPI]{coherent processing interval}
\acrodef{cs}[CS]{compressed sensing}
\acrodef{tbd}[TBD]{track-before-detect}
\acrodef{snr}[SNR]{signal-to-noise ratio}
\acrodef{ula}[ULA]{uniform linear array}
\acrodef{doa}[DoA]{direction of arrival}
\acrodef{pri}[PRI]{pulse repetition interval}
\acrodef{ctft}[CTFT]{continuous-time Fourier transform}
\acrodef{lasso}[LASSO]{least absolute shrinkage and selection operator}
\acrodef{awgn}[AWGN]{additive white Gaussian noise}
\acrodef{cwna}[CWNA]{continuous white noise acceleration}
\acrodef{ista}[ISTA]{iterative shrinkage-thresholding algorithm}
\acrodef{fista}[FISTA]{fast iterative shrinkage-thresholding algorithm}
\acrodef{dft}[DFT]{discrete Fourier transform}
\acrodef{gmm}[GMM]{Gaussian mixture model}
\acrodef{mf}[MF]{matched filtering}
\acrodef{mmv}[MMV]{multiple measurement vectors}
\acrodef{mimo}[MIMO]{multiple-input-multiple-output}
\acrodef{sbl}[SBL]{sparse Bayesian learning}
\acrodef{map}[MAP]{maximum a posteriori}
\acrodef{iht}[IHT]{iterative hard thresholding}
\acrodef{amp}[AMP]{approximate message passing}
\acrodef{rs}[RS]{replica symmetric}
\acrodef{tap}[TAP]{Thouless-Anderson-Palmer}
\acrodef{eccm}[ECCM]{electronic counter-countermeasures}
\acrodef{camp}[CAMP]{complex Approximate Message Passing}
\acrodef{glrt}[GLRT]{generalized likelihood ratio test}
\acrodef{mle}[MLE]{maximum likelihood estimation}
\acrodef{camp}[CAMP]{complex approximate message passing}
\acrodef{ecdf}[ECDF]{empirical cumulative distribution function}
\acrodef{crod}[CROD]{Complex Row-Orthogonal Debiased detector}
\acrodef{rod}[ROD]{Row-Orthogonal Debiased detector}
\acrodef{dct}[DCT]{discrete cosine transformation}
\acrodef{ks}[KS]{Kolmogorov-Smirnov}
\acrodef{dwld}[DWLD]{the debiased weighted LASSO detector}
\acrodef{rip}[RIP]{restricted isometry property}
\acrodef{nwld}[NWLD]{the naive weighted LASSO detector}
\acrodef{dld}[DLD]{the debiased LASSO detector}

\begin{document}

	\title{Compressed Sensing Radar Detectors\\ based on Weighted LASSO}
	
	\author{\IEEEauthorblockN{Siqi Na\IEEEauthorrefmark{1},
    Yoshiyuki Kabashima\IEEEauthorrefmark{2},
    Takashi Takahashi\IEEEauthorrefmark{2},
    Tianyao Huang\IEEEauthorrefmark{1},
    Yimin Liu\IEEEauthorrefmark{1} and
    Xiqin Wang\IEEEauthorrefmark{1}}
    
    \IEEEauthorblockA{Email: nsq17@mails.tsinghua.edu.cn kaba@phys.s.u-tokyo.ac.jp takashi-takahashi@g.ecc.u-tokyo.ac.jp \{huangtianyao, yiminliu, wangxq\_ee\}@tsinghua.edu.cn
    \\}
    \IEEEauthorblockA{\IEEEauthorrefmark{1}
    Department of Electronic Engineering, Tsinghua University, Beijing, China
    \\}
    \IEEEauthorblockA{\IEEEauthorrefmark{2}
    Institute for Physics of Intelligence, Department of Physics Graduate School of Science, The University of Tokyo, Japan}
    \thanks{The work of Tianyao Huang was supported by the National Natural Science Foundation of China under Grant 62171259. }
    }

	\maketitle
	
	\begin{abstract}
	    The compressed sensing (CS) model can represent the signal recovery process of a large number of radar systems. 
	    The detection problem of such radar systems has been studied in many pieces of literature through the technology of debiased \emph{least absolute shrinkage and selection operator (LASSO)}. 
	    While naive LASSO treats all the entries equally, there are many applications in which prior information varies depending on each entry. 
	    Weighted LASSO, in which the weights of the regularization terms are tuned depending on the entry-dependent prior, is proven to be more effective with the prior information by many researchers. 
	    In the present paper, existing results obtained by methods of statistical mechanics are utilized to derive the debiased weighted LASSO estimator for randomly constructed row-orthogonal measurement matrices.  
	    Based on this estimator, we construct a detector, termed the debiased weighted LASSO detector (DWLD), for CS radar systems and prove its advantages. 
	    The threshold of this detector can be calculated by false alarm rate, which yields better detection performance than naive weighted LASSO detector (NWLD) under the Neyman-Pearson principle. 
	    The improvement of the detection performance brought by tuning weights is demonstrated by numerical experiments. 
	    With the same false alarm rate, the detection probability of DWLD is obviously higher than those of NWLD and the debiased (non-weighted) LASSO detector (DLD). 
    \end{abstract}
    
    
    \IEEEpeerreviewmaketitle

	\section{Introduction}
    \label{sec:intro}
    
    
    The signal recovery process of radar systems can be represented by a simple linear regression model as follow 
    \begin{equation}
	    \label{eq:csmodel}
	    {\bm{y}} = {\bm{A}}{\bm{x}}_0 + {\bm{\xi}},
	\end{equation}
	where ${\bm{A}} \in {\mathbb{C}}^{M\times N}$ is called measurement matrix, which contains information of the radar transmit waveform. 
	Noise vector ${\bm{\xi}} \in {\mathbb{C}}^M$ is usually regarded as \ac{awgn} with i.i.d. components $\xi_i \sim {\cal {CN}}\left( {0,\sigma^2} \right)$. 
	Usually, one aims to recover ${\bm{x}}_0 \in {\mathbb{C}}^N$, which contains information about the radar observation scene, from the received signal ${\bm{y}} \in {\mathbb{C}}^M$. 
	The increasing development of \ac{cs} technology in recent years has resulted in many \ac{cs} radar systems, in which signal detection from fewer measurements ($M<N$) is required. 
	In addition, we find that a number of radar emission waveforms conform to the row-orthogonal assumption, such as partial observation of a pulse Doppler radar system \cite{xiao2018distributed}, frequency agile radar system \cite{huang2018analysis, wang2021randomized}, sub-Nyquist radar systems \cite{cohen2018sub, na2018tendsur}. 
	We here defined matrix $A$ as ``row-orthogonal" when it is constructed randomly so as to satisfy $A_{i \cdot} A_{j \cdot}^H = \delta_{ij}$ $1\le i, j \le M$, where $\delta_{ij} = 1$ for $i = j$ and $0$, otherwise. 
	
	Based on results in literature, if ${\bm{x}}_0$ is sparse enough, it can be precisely recovered from ${\bm{y}}$ in a noise-free scene. 
	That is, the upper bound of $k$, which is the number of non-zero entries in ${\bm{x}}_0$, for the perfect recovery is strictly limited by some properties of measurement matrix ${\bm{A}}$ such as the {\emph{compression rate}} $\gamma = M / N$. 
	To solve the underdetermined recovery problem \eqref{eq:csmodel}, \ac{cs} methods are usually applied, in which the \ac{lasso} \cite{tibshirani1996regression} is a frequently used technique whose behavior has been studied in much literature \cite{greenshtein2004persistence, bickel2009simultaneous, candes2007dantzig, raskutti2011minimax, meinshausen2006high, zhao2006model, wainwright2009sharp}. 
	
	The complex-valued \ac{lasso} estimator $\hat {\bm{x}}^{\rm{LASSO}}$ is given by
	\begin{equation}
	    \hat {\bm{x}}^{\rm{LASSO}} = \mathop {\arg \min} \limits_{{\bm{x}}} \left\{ \frac{1}{2} {\left\| {\bm y} - {\bm A}{\bm x} \right\|_2^2} +  \lambda {\left\|{\bm x} \right\|_1} \right\},
	\end{equation}
	where 
	\begin{equation}
	    {\left\|{\bm x} \right\|_1} \buildrel \textstyle. \over = \sum\limits_{i = 1}^{N} {\sqrt{ \left( {\rm{Re}}\left( x_i \right) \right)^2 + \left( {\rm{Im}}\left( x_i \right) \right)^2}}.
	\end{equation}
	Based on debiased \ac{lasso}, which is also known as \emph{desparsified \ac{lasso}} in the context of statistics, given by
	\begin{equation}
        \label{eq:debiased_LASSO}
        {\hat {\bm x}}^{\rm {d}} = {\hat {\bm x}}^{\rm {LASSO}} + \frac{1}{\Lambda} {\bm A}^H ({\bm y} - {\bm A} {\hat {\bm x}}^{\rm {LASSO}}),
    \end{equation}
    where $\Lambda > 0$ is the {\emph{debiased coefficient}} computed from known variables, there are also some researches \cite{javanmard2014hypothesis, anitori2012design, takahashi2018statistical, na2022compressed} on the following detection problems
	\begin{equation}
        \label{eq:hypothesis_testing}
        \left\{ {\begin{array}{*{20}{l}}
        {{{\cal H}_{0, i}}:{x_{0,i}} = 0}, \\
        {{{\cal H}_{1, i}}:{x_{0,i}} \ne 0},
        \end{array}} \right.
    \end{equation}
    for $i = 1, 2, \ldots, N$. 
	As mentioned above, the indices of the non-zero elements in ${\bm{x}}_0$ indicate information about the target, and determining whether each element in ${\bm{x}}_0$ is non-zero or not can inform us about the existence and location of the targets. 
	Consequently, dealing with problems \eqref{eq:hypothesis_testing} means to detect targets element-wisely, which is very critical for modern radar systems. 
    
    There are also some applications for radar systems that we may acquire some prior information about ${\bm{x}}_0$, which refers to the support set or the probability that each element in ${\bm{x}}_0$ is non-zero. 
    For instance, the multi-frame observations with tracking \cite{na2020track} brings us the prediction of the target's position in the next frame. 
    A weighted version of \ac{lasso}, weighted \ac{lasso} (or weighted $\ell_1$ minimization for noise-free scenario), has been studied to exploit prior information in the process of signal recovery \cite{vaswani2010modified, liang2010compressed, needell2017weighted, lian2018weighted}, which is given by
    \begin{equation}
        \label{eq:WL}
	    \hat {\bm{x}}^{\rm{WL}} = \mathop {\arg \min} \limits_{{\bm{x}}} \left\{ \frac{1}{2} {\left\| {\bm y} - {\bm A}{\bm x} \right\|_2^2} +  \sum\limits_{i = 1}^{N} \lambda_i \| x_i \| \right\}, 
	\end{equation}
	where $ {\bm \lambda} \buildrel \textstyle. \over = [\lambda_1, \lambda_2, \ldots, \lambda_N ]^T$ is the vector of weights. 
	Researches mentioned above are mainly based on \ac{rip} of measurement matrix $\bm A$ and provide results on the upper bound of $\| \hat {\bm{x}}^{\rm{WL}} - {\bm{x}}_0 \|_2$. 
	Reference \cite{tanaka2010optimal} shows the advantages of proper weighting according to prior information for noise-free cases. 
	
	As for the debiased strategy, we are interested in the debiased weighted \ac{lasso} estimator, given by
	\begin{equation}
        \label{eq:debiased_WL}
        {\hat {\bm x}}^{\rm {d}} = {\hat {\bm x}}^{\rm {WL}} + \frac{1}{\Lambda} {\bm A}^H ({\bm y} - {\bm A} {\hat {\bm x}}^{\rm {WL}}).  
    \end{equation}
    With results in \cite{celentano2020lasso} and some variable substitution, one can obtain a debiased weighted \ac{lasso} estimator under random Gaussian design, that is, the elements of  $\bm A$ are i.i.d. drawn from a Gaussian distribution. 
    Our goal is to derive a debiased weighted \ac{lasso} estimator, particularly for complex-valued row-orthogonal measurement matrix, and construct its related detector, which we term \ac{dwld}. 
    
    In this paper, we provide the framework of \ac{dwld} and analyze its detection performance by comparing \ac{dwld} with \ac{nwld}, which is defined as a set of detectors rejecting ${\cal H}_{0, i}$ if $|{\hat x}^{\rm{WL}}_i|$ is greater than a given threshold. 
    When the measurement matrix is complex-valued row-orthogonal, the concrete implementation of \ac{dwld} is presented employing results obtained by methods of statistical mechanics \cite{takahashi2018statistical, na2022compressed}. 
    Merits of \ac{dwld} are twofold: the threshold of \ac{dwld} can be analytically calculated by the probability of false alarm, which enables \ac{dwld} to achieve a better detection performance than \ac{nwld} under the Neyman-Pearson principle. 
    We propose a method to optimize the setting of weights ${\bm \lambda}$ with the goal of improving the detection performance. 
    We also demonstrate that proper weighting can improve detection performance than \ac{dld} \cite{na2022compressed} through numerical experiments.
	
	The present paper is organized as follows. 
	In Section \ref{sec:method}, we implement \ac{dwld} for compressed sensing radar system with prior information providing theoretical consideration on its detection performance. 
	The approach to optimize the weights, in order to improve detection performance, is also proposed. 
	Section \ref{sec:result} provides some numerical results of the detection performance of \ac{dwld}. 
	We conclude the paper in Section \ref{sec:conclusion}. 
	
	Throughout the paper, we use $a$, $\bm a$ and $\bm A$ as a number, a vector and a matrix, respectively. 
	For a set $S$, $\# S$ denotes its cardinality. 
	Function $\delta(\cdot)$ is the Dirac's delta function and $\Theta (x)$ is Heaviside's step function.
	The operators $(\cdot)^T$, $(\cdot)^*$ and $(\cdot)^H$ represent the transpose, conjugate and conjugate transpose of a component, respectively. 
	The indicator function is of a subset $A$ of a set $X$ is presented by ${\bm 1}_A (x)$. 
	
    \section{Design and analysis of debiased weighted LASSO detector}
    \label{sec:method}
    
    In this section, we first provide some metrics with regard to detection performance in Section \ref{subsec:preliminaries}. 
    Then the framework of \ac{dwld} together with \ac{nwld} is introduced in Section \ref{subsec:design}.
    We also give a concrete implementation of \ac{dwld} under row-orthogonal design, including the procedure for calculating thresholds based on false alarm rates. 
    The advantages of \ac{dwld} are elaborated mainly by comparing with \ac{nwld} in Section \ref{subsec:analysis}. 
    At last, the method to optimize the weights $\bm \lambda$ is proposed in Section \ref{subsec:opt}. 
    	
	\subsection{Preliminaries}
	\label{subsec:preliminaries}
	
	    For the $N$ detection problems described in \eqref{eq:hypothesis_testing}, we here define the following performance metrics.
    	Suppose we perform the tests for $T$ trials. 
	    Denote by
	    \begin{equation}
	        {\varphi_i^{(t)}} = \left\{ {\begin{array}{*{20}{c}}
            {1,}&{{\text{if detector reject }}{{\cal{H}}_{0,i}}{\text{ in $t$-th trial}};}\\
            {0,}&{{\text{otherwise,}}}
            \end{array}} \right.
	    \end{equation}
	    for $i = 1, \ldots, N$ and $t = 1, \ldots, T$. 
	    Define the probability of false alarm  $P_{fa, i}$ for the $i$-th problem or entry of ${\bm x}_0$ as
	    \begin{equation}
	        P_{fa, i} = \mathop {\lim }\limits_{T \to \infty } \frac{\sum\limits_{t = 1}^T { \varphi_{i}^{(t)} \cdot {\bm 1}_{S_t^c}(i) }}{\sum\limits_{t = 1}^T { {\bm 1}_{S_t^c}(i) }} ,
	    \end{equation}
	    and the probability of detection $P_{d, i}$ as
	    \begin{equation}
	        P_{d, i} = \mathop {\lim }\limits_{T \to \infty } \frac{\sum\limits_{t = 1}^T { \varphi_{i}^{(t)} \cdot {\bm 1}_{S_t}(i) }}{\sum\limits_{t = 1}^T { {\bm 1}_{S_t}(i) }} ,
	    \end{equation}
	    where $S_t = \left\{ i: x_{0, i}^{(t)} \ne 0 \right\}$ is the support set of ${\bm x}_0^{(t)}$ with $S_t^c = \{1, \ldots, N\} \backslash S_t$.
	    
	    The prior information ${\bm p} = [p_1, p_2, \ldots, p_N]^T$ of ${\bm x}_0$ is defined as the probability that each entry of ${\bm x}_0$ is non-zero, such that
        \begin{equation}
            P({\bm{x}}_0) = \prod\limits_{i = 1}^N \left[  (1 - p_i) \delta(x_{0, i}) + p_i Q_i(x_{0, i}) \right], 
        \end{equation}
        where $Q_i(x_{0, i})$ is the distribution of non-zero entry of ${\bm x}_0$.
        
        Since $N$ detection problems are processed simultaneously, we define the total false alarm rate $\bar P_{fa}$ as follow,  
        \begin{equation}
            \label{eq:bar_P_fa}
            {\bar P_{fa}} = \mathop {\lim }\limits_{T \to \infty } \frac{\sum\limits_{t = 1}^T \sum\limits_{i = 1}^N { \varphi_{i}^{(t)} \cdot {\bm 1}_{S_t^c}(i) }}{\sum\limits_{t = 1}^T \sum\limits_{i = 1}^N { {\bm 1}_{S_t^c}(i) }},  
        \end{equation}
        and similarly the total detection rate $\bar P_d$
        \begin{equation}
            \label{eq:bar_P_d}
            {\bar P_d} = \mathop {\lim }\limits_{T \to \infty } \frac{\sum\limits_{t = 1}^T \sum\limits_{i = 1}^N { \varphi_{i}^{(t)} \cdot {\bm 1}_{S_t}(i) }}{\sum\limits_{t = 1}^T \sum\limits_{i = 1}^N { {\bm 1}_{S_t}(i) }}.  
        \end{equation}
        With the definition of prior information $\bm p$ above, one can say that
        \begin{equation}
            p_i = \mathop {\lim }\limits_{T \to \infty } \frac{1}{T} \sum\limits_{t = 1}^T { {\bm 1}_{S_t}(i) }.
        \end{equation}
        Therefore, 
        \begin{eqnarray}
            &&{\bar P_{fa}} = \sum\limits_{i = 1}^N (1-p_i) P_{fa, i} / \sum\limits_{i = 1}^N (1-p_i), \\
            &&{\bar P_{d}} = \sum\limits_{i = 1}^N p_i P_{d, i} / \sum\limits_{i = 1}^N p_i. 
        \end{eqnarray}
        Note that if we take the limit $N \to \infty$, these definition will be the same as that in \cite{na2022compressed}, with \eqref{eq:bar_P_fa} and \eqref{eq:bar_P_d} considering prior information. 
	
	\subsection{Design of the debiased weighted LASSO detector}
	\label{subsec:design}
	    
    	\begin{figure*}
    	    \centering
    	    \includegraphics[width=15cm]{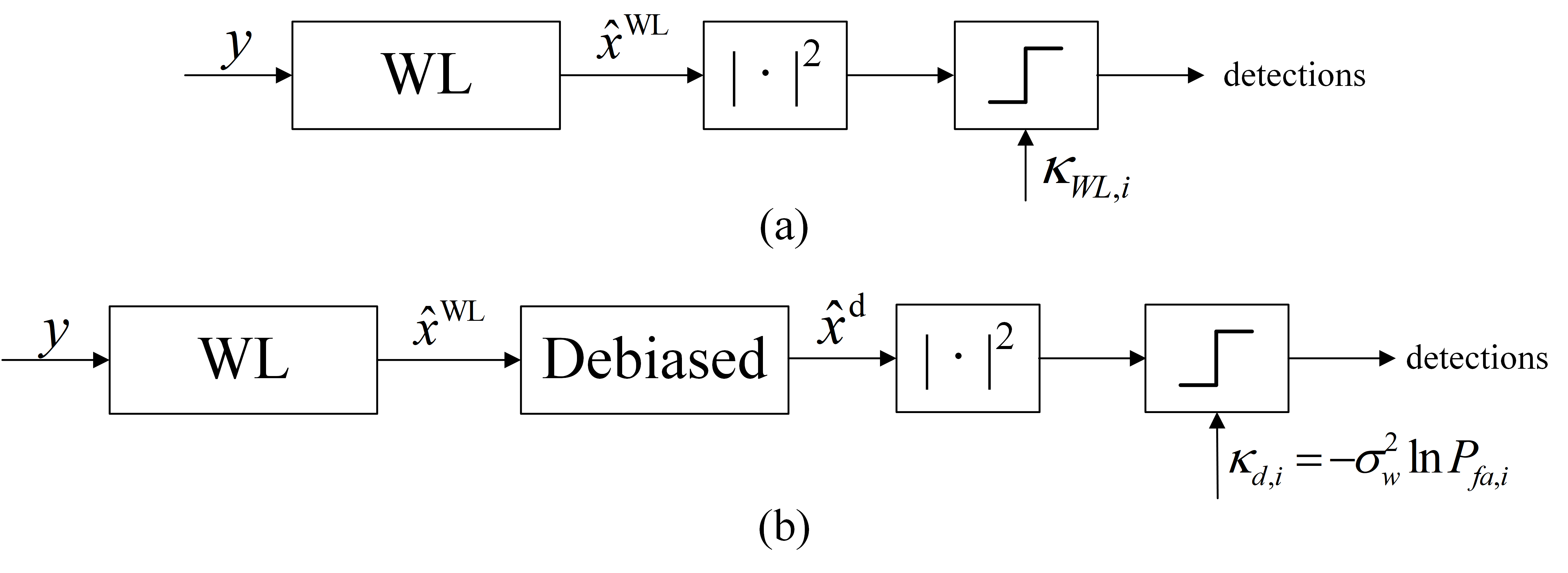}
    	    \caption{Frameworks of two detectors for detection problem \eqref{eq:hypothesis_testing}: (a) naive weighted \ac{lasso} detector (NWLD); (b) debiased weighted \ac{lasso} detector (DWLD). }
    	    \label{fig:Debiased_weighted_LASSO_detector_framework}
    	\end{figure*}
    	
    	Similarly to \cite{na2022compressed}, we demonstrate and compare \ac{dwld} with \ac{nwld}. 
    	Their frameworks are shown in Fig. \ref{fig:Debiased_weighted_LASSO_detector_framework} (a) and (b). 
    	The main difference between the two detectors is in the test statistics and thresholds. 
    	The latter performs a ``debiased" procedure, which is a linear transform of the weighted \ac{lasso} estimator, and applies calculable thresholds with given false alarm rate for each entry. 
    	The details of the superiority of \ac{dwld} and their proofs will be presented in Sec. \ref{subsec:analysis}. 
    	
    	Due to the random nature of the row-orthogonal matrix ${\bm A}$, ${\hat {\bm x}}^{\rm {d}}$ is expected to follow Gaussian distribution with mean ${\bm x}_0$. 
    	Actually, this is the case as long as ${\bm A}$ is ``right rotation invariant", which means that the right singular basis of ${\bm A}$ is regarded as a random sample from the Haar measure of $N \times N$ orthogonal (or unitary) matrices. 
    	Further, in such cases, methods of statistical mechanics offer formulas to appropriately construct ${\hat {\bm x}}^{\rm {d}}$ from the (weighted) \ac{lasso} estimator and to analytically evaluate the variance of ${\hat {\bm x}}^{\rm {d}}$ depending on the asymptotic eigenvalue distribution $\rho_{\bm J}(s)$ of ${\bm J} = {\bm A}^T {\bm A}$ (or ${\bm J} = {\bm A}^H {\bm A}$) \cite{takahashi2018statistical, na2022compressed}. 
    	Utilizing the formulas, we here propose \ac{dwld} for complex-valued row-orthogonal observation matrix, whose procedure is shown in Algorithm \ref{alg:DWLD}. 
    	Due to space limitation, we leave its derivation to \cite{na2022compressed}.  
    	
    	\begin{algorithm}
            \caption{\ac{dwld} under complex row-orthogonal design}
            \label{alg:DWLD}
            \begin{algorithmic}[1]
                \Require $\bm y$, $\bm A$, weights ${\bm \lambda}$, probability of false alarm $\{P_{fa, i}\}_{i = 1}^N$, variance of noise $\sigma^2$
                \Ensure debiased weighted \ac{lasso} estimator ${\hat {\bm x}}^{\rm {d}}$, threshold $\{\kappa_{d, i}\}_{i = 1}^N$
                \State Let 
                    \begin{equation}
                	    \hat {\bm{x}}^{\rm{WL}} = \mathop {\arg \min} \limits_{{\bm{x}}} \left\{ \frac{1}{2} {\left\| {\bm y} - {\bm A}{\bm x} \right\|_2^2} +  \sum\limits_{i = 1}^{N} \lambda_i \| x_i \| \right\}.
                	    \nonumber
                	\end{equation}
            	\State The debiased weighted \ac{lasso} estimator is obtained from
                    \begin{equation}
                	    \label{eq:debiased_WL_RO}
                        {\hat {\bm x}}^{\rm {d}} = {\hat {\bm x}}^{\rm {WL}} + \frac{1}{\Lambda_{\rm{CRO}}} {\bm A}^H ({\bm y} - {\bm A} {\hat {\bm x}}^{\rm {WL}}),
            	    \end{equation}
            	    with $\Lambda_{\rm{CRO}}$ and $\rho_{\rm{CA}}$ are solved from the following fixed point equation
                    \begin{eqnarray}
                        \Lambda_{\rm{CRO}} &=& \frac{\gamma - \rho_{\rm{CA}}}{1 - \rho_{\rm{CA}}}, \nonumber \\
                        \label{eq:alg_rho_CA}
            	        \rho_{\rm{CA}} &=& \frac{1}{2N} \sum \limits_{i = 1}^N \left[ \left( 2 - \frac{\lambda_i}{\Lambda_{\rm{CRO}} \left| {\hat x_i^{\rm{WL}}} \right| + \lambda_i } \right) \nonumber \right. \\
            	        && \left. \cdot  \Theta \left( \left| {\hat x_i^{\rm{WL}}} \right| \right) \right] . 
                    \end{eqnarray}
                \State The threshold $\{\kappa_{d, i}\}_{i = 1}^N$ is given by
                    \begin{equation}
                        \kappa_{d, i} = -\sigma_w^2 \ln{P_{fa, i}}, \nonumber
                    \end{equation}
                    where
        	        \begin{eqnarray}
        	            \label{eq:alg_sigma}
                        &&\sigma_w^2 = \frac{\gamma (1-\gamma)}{\left( \gamma - \rho_{\rm{CA}} \right)^2} \overline{\rm{RSS}} + \sigma^2, \\
                        \label{eq:alg_RSS}
                        &&\overline{\rm{RSS}} = \frac{1}{M}  \left\| \bm{y} - \bm{A} {\hat {\bm x}}^{\rm {WL}} \right\|_2^2.  
                    \end{eqnarray}
                
            \end{algorithmic}
        \end{algorithm}
        
        One can deduce results similar to \eqref{eq:debiased_WL_RO} for random Gaussian measurement matrices from some conclusions in \cite{celentano2020lasso}. 
    	More specifically, with Theorem 10 in \cite{celentano2020lasso}, the debiased coefficient for random Gaussian design is $\Lambda_G = \gamma - \rho_a$, where $\rho_a = \# \{ i | {\hat x_i^{\rm{WL}}} \ne 0 \}/N$ denotes the {\emph{active component density}} for real-valued situation. 
    	We emphasize that the methodologies in \cite{takahashi2018statistical, na2022compressed} not only reproduce the same result for the Gaussian matrices but also can offer the debiased coefficients for general right rotation invariant matrices. 
        
    \subsection{Analysis of detection performance}
	\label{subsec:analysis}
	
	In general, the advantage of \ac{dwld} is that its threshold $\kappa_{d, i}$ can be calculated analytically from the false alarm rate $P_{fa, i}$.  
	On the other hand, due to the fact that there is no analytic solution for weighted \ac{lasso}, one cannot establish the relationship between the threshold $\kappa_{WL, i}$ of \ac{nwld} and the false alarm rate $P_{fa, i}$. 
	We will next introduce the two benefits brought by \ac{dwld} in Section \ref{subsubsec:analytical} and \ref{subsubsec:detection_performance}, respectively. 
    	
    	\subsubsection{Analytical expression of false alarm probability}
    	\label{subsubsec:analytical}
    	
    	Controlling the probability of false alarm is very important for radar applications. 
        Under the Neyman-Pearson criterion, the performance of the detectors can be compared only if the upper bound of the false alarm rate is controlled. 
        Therefore, we state that the most significant strength of \ac{dwld} is that the analytical relationship between the threshold $\kappa_{d, i}$ and the false alarm rate $P_{fa, i}$ is available.
        As shown in \cite{celentano2020lasso}, for the debiased estimator \eqref{eq:debiased_WL}, the empirical distribution of the difference
        \begin{equation}
            \label{eq:w}
            {\bm{w}} \buildrel \textstyle. \over = {\hat {\bm x}}^{\rm {d}} - {\bm{x}}_0,  
        \end{equation}
        converges weakly to a zero-mean Gaussian distribution as $N \to \infty$ for a specific $\Lambda$.
	    Denote the sample variance of ${\bm{w}}$ by $\sigma_w^2$, one can get the following analytical relationship between the probability of false alarm $P_{fa, i}$ and the threshold $\kappa_{d, i}$ of the detector. 
	    
	    \begin{theorem}
	        \label{theorem:pfa_kappa}
	        When $N \to \infty$, the false alarm probability of $i$-th entry $P_{fa, i}$ of \ac{dwld} satisfies: 
	        \begin{equation}
	            \label{eq:pfa_kappa}
	            \kappa_{d, i} = - \sigma_w^2 \ln P_{fa, i},
	        \end{equation}
	        where the test is
	        \begin{equation}
    	        {\varphi_{i}} = \left\{ {\begin{array}{*{20}{l}}
                {1,}&{\left| \hat {x}^{\rm{d}}_i \right|^2 > \kappa_{d, i} ;}\\
                {0,}&{{\text{otherwise.}}}
                \end{array}} \right.
    	    \end{equation}
	    \end{theorem}
	    
	    The readers are referred to \cite{na2022compressed} for the proof. 
	    Due to the fact that almost all of the solutions obtained by \ac{cs} methods does not have a closed form, neither the distribution of the solutions, it is hard to obtain such conclusion for conventional \ac{cs} detectors. 
	    
	    \subsubsection{Better detection performance}
    	\label{subsubsec:detection_performance}
	    
	    Based on the definition and subdifferential properties of a convex function, one can draw the following conclusion. 
	    
	    \begin{theorem}
	    
	        \label{theorem:detection_performance}
	        
	        For the same false alarm probability $P_{fa, i}$, the detection probability of \ac{dwld} $P_{d, i}^{(1)}$ is not less than that of \ac{nwld} $P_{d, i}^{(2)}$.

	    \end{theorem}
	    
	    The proof of this conclusion is similar as that of Theorem 2.7 in \cite{na2022compressed}, which we refer the readers to. 
	    Theorem \ref{theorem:detection_performance} guarantees that \ac{dwld} has a better detection performance comparing with weighted \ac{lasso} detector.

    \subsection{Optimization of weights}
    \label{subsec:opt}
    
        One can easily conclude from the previous analysis that the smaller $\sigma_w^2$ is, the better the detection performance of \ac{dwld} can achieve. 
        Given other parameters (e.g. distribution and size of measurement matrix ${\bm A}$, variance of the noise $\sigma^2$, distribution of ${\bm x}_0$, etc.), $\sigma_w^2$ can be expressed as a function of weights $\bm \lambda$ and prior information $\bm p$, given by
        \begin{equation}
            \sigma_w^2 = f_1 \left( {\bm \lambda}, {\bm p} \right). 
        \end{equation}
        However, $f_1$ is unavailable because we do not know ${\bm x}_0$. 
        Instead, we can obtain its estimation $\hat \sigma_w^2$ by the method developed in \cite{takahashi2018statistical} and \cite{na2022compressed}, which can be represented by
        \begin{equation}
            {\mathbb{E}}_{{\bm A}, {\bm \xi}, {\bm x}_0} \left( \hat \sigma_w^2 \right) = f_2 \left( {\bm \lambda}, {\bm p} \right), 
        \end{equation}
        where the specific implementation of $f_2$ can be obtained from Algorithm \ref{alg:DWLD}. 
        Therefore, we can find the optimal weights $\bm \lambda$ for the prior information ${\bm p}$ by optimizing $f_2$, given by
        \begin{equation}
            \label{eq:opt_lambda}
            {\bm \lambda}^* = \mathop {\arg \min} \limits_{{\bm \lambda}} \left\{ f_2 \left( {\bm \lambda}, {\bm p} \right) \right\} \buildrel \textstyle. \over = f_3({\bm p}).
        \end{equation}
        
        Unfortunately, $f_2$ dose not have an analytic expression. 
        Employing heuristic algorithms to solve an $N$-dimensional parameter optimization problem is not very practical. 
        Therefore, we reduced the degrees of freedom by parameterizing the weight vector $\bm{\lambda}$ with a few hyper-parameters, such as a linear model $\lambda^*_i \simeq \lambda_0 - \alpha p_i$, or an exponential model $\lambda^*_i \simeq \lambda_0 / p_i^{\alpha}$, and obtained sub-optimal results.  
        Consequently, considering the linear model for example, the optimization problem in \eqref{eq:opt_lambda} becomes
        \begin{equation}
    	    \label{eq:opt_lambda2}
	        \{\lambda_0, \alpha\} = \mathop {\arg \min} \limits_{\{a, b \}} \left\{ f_2 \left( \{a - b p_i\}_{i = 1}^N, {\bm p} \right) \right\}. 
    	\end{equation}
    	The simulation results in Section \ref{sec:result} will verify the effectiveness of proposed weights optimization approach.

    \section{Numerical Experiments}
    \label{sec:result}
    
    We will demonstrate the detection performance of \ac{dwld} through numerical simulation, mainly under row-orthogonal design. 
    We first verify the ability of \ac{dwld} to determine the threshold based on the false alarm rate in Sec. \ref{subsec:comparison}, in which we compare \ac{dwld} and \ac{nwld}. 
    Then, in Sec. \ref{subsec:detection_performance}, the comparison between \ac{dwld} with \ac{dld} under complex row-orthogonal design, which is termed CROD in \cite{na2022compressed}, illustrates that proper setting of weights $\bm \lambda$ can lead to improved detection performance. 
        
        \subsection{Settings}
        \label{subsec:settings}
        
            In all the numerical experiments, we artificially generate the original signal ${\bm{x}}_0$, observation matrix ${\bm{A}}$ and the noise ${\bm{\xi}}$. 
            The original signal ${\bm{x}}_0$ is generated from the Bernoulli-Gaussian distribution and the entries are independent with prior information ${\bm p}$, thus
            \begin{equation}
                P({\bm{x}}_0) = \prod\limits_{i = 1}^N \left[  (1 - p_i) \delta(x_{0, i}) + \frac{p_i}{\pi \sigma_x^2} {\rm{e}}^{-|x_{0, i}|^2 / \sigma_x^2} \right].
            \end{equation}
            The measurement matrix ${\bm{A}}$ is decided by the radar system. 
            The entries of the noise ${\bm{\xi}}$ are i.i.d. complex Gaussian variables: $\xi_i \sim {\cal{CN}}(0, \sigma^2)$. 
            We adopt the \ac{mf} definition of the \ac{snr} \cite{na2018tendsur}, such that
            \begin{equation}
                {\rm{SNR}} = \frac{\gamma \sigma_x^2}{\sigma^2}.
            \end{equation}

        \subsection{Comparison of \ac{dwld} and \ac{nwld}}
		\label{subsec:comparison}
		
			\begin{figure}
				\centering
				\includegraphics[width=7.5cm]{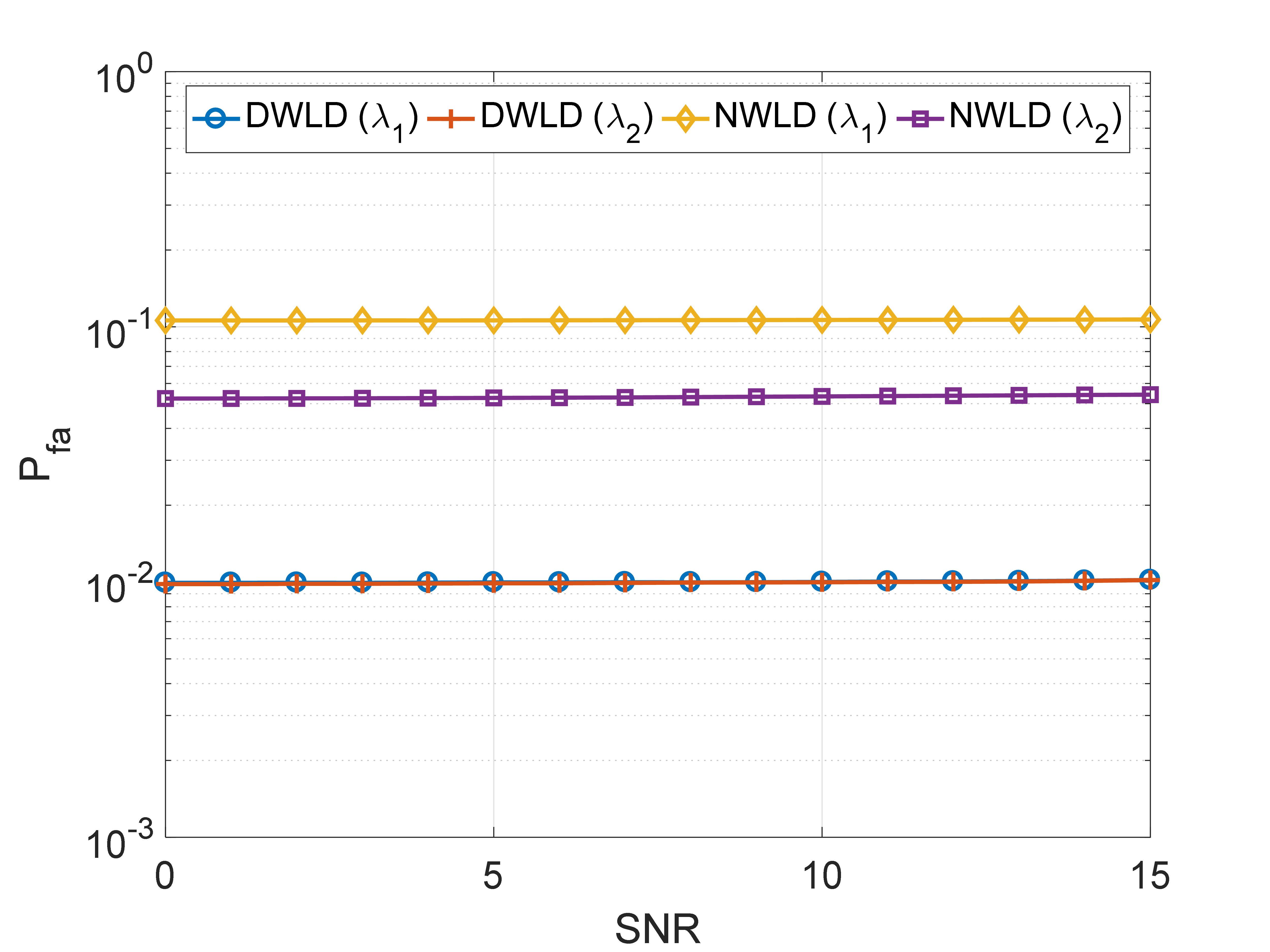}
				\caption{Total false alarm rate of \ac{dwld} and \ac{nwld} vary with SNR for the first prior information setting \eqref{eq:pi1}. Here $\gamma = 0.5$.}
				\label{Fig:DNWLD_Pfa_Pd_PF_SNR1}
			\end{figure}
			
			\begin{figure}
				\centering
				\includegraphics[width=7.5cm]{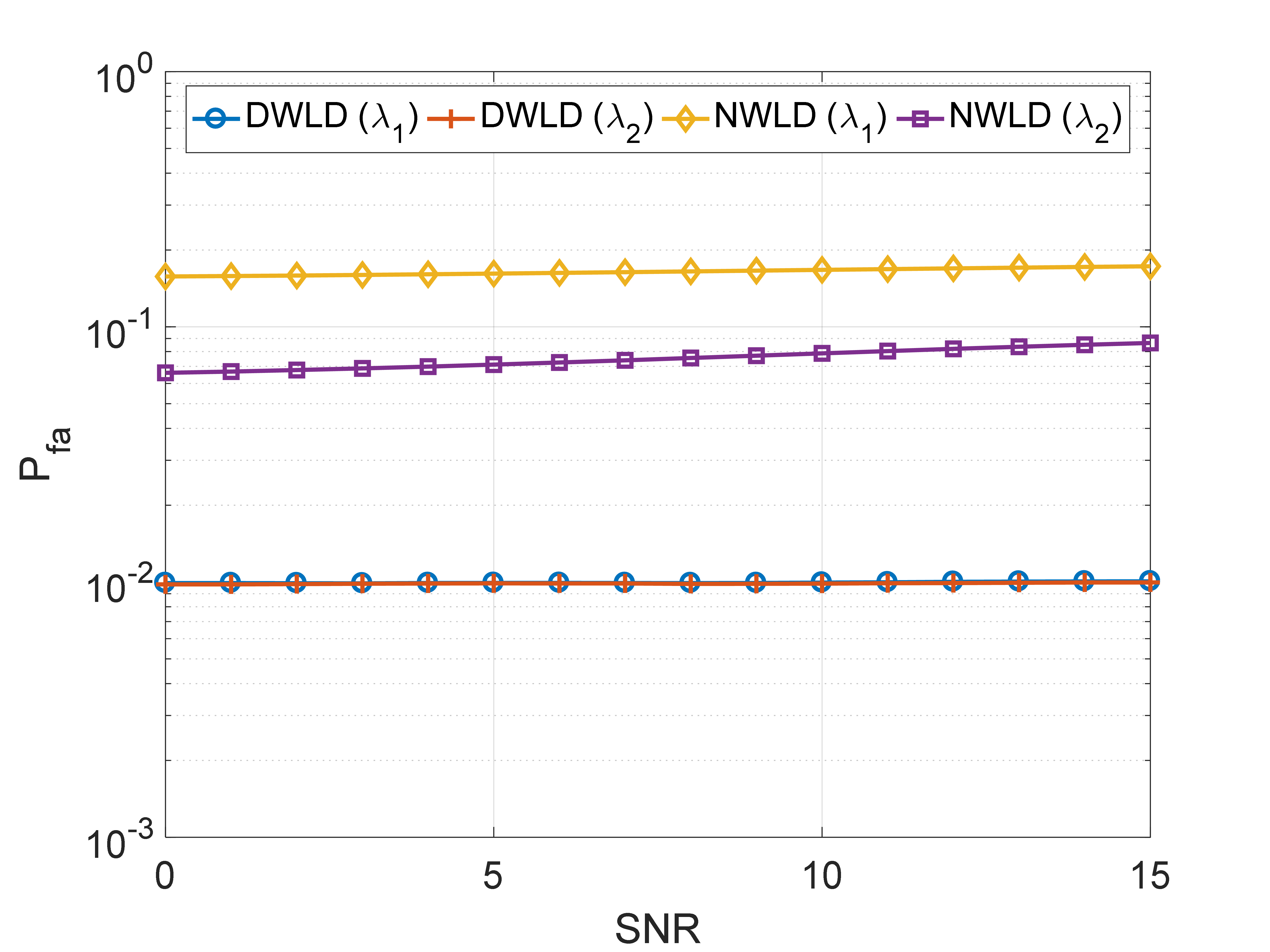}
				\caption{Total false alarm rate of \ac{dwld} and \ac{nwld} vary with SNR for the second prior information setting \eqref{eq:pi2}. Here $\gamma = 0.5$.}
				\label{Fig:DNWLD_Pfa_Pd_PF_SNR2}
			\end{figure}
            
            We apply \ac{dwld} and \ac{nwld} for solving the sub-Nyquist radar detection problem, in which the observation matrix $\bm A$ is partial Fourier, and examine their false alarm rate. 
            In the following experiments, we set the length of ${\bm x}_0$ to $N = 512$ and the variance of noise $\sigma^2$ to $0.01$. 
			The probability of false alarm is set to $0.01$. 
			All the results are obtained by $10^4$ Monte-Carol trials. 
            
            The result of the first experiment is shown in Fig. \ref{Fig:DNWLD_Pfa_Pd_PF_SNR1}, where the prior information $\bm p$ is set as
			\begin{equation}
			    \label{eq:pi1}
			    {p_i} = \left\{ {\begin{array}{*{20}{c}}
                {0.01,}&{1 \le i \le \frac{3}{4}N,}\\
                {0.8,}&{\frac{3}{4}N < i \le N.}
                \end{array}} \right.
			\end{equation}
			The weight vectors ${\bm \lambda}_1$ and ${\bm \lambda}_2$ of the two detectors are set to $\lambda_{1, i} = 0.1 - 0.1 p_i$ and $\lambda_{2, i} = 0.15 - 0.15 p_i$, respectively. 
			The threshold of \ac{nwld} is set to $0$, which corresponds to the naive decision based on the weighted LASSO estimator. 
			From the result we can see that \ac{dwld} has a good ability to control the false alarm rate when different parameters vary while \ac{nwld} dose not. 
			
			The second one sets the prior information $\bm p$ as
			\begin{equation}
			    \label{eq:pi2}
			    {p_i} = \left\{ {\begin{array}{*{20}{c}}
                {0.05,}&{1 \le i \le \frac{3}{4}N,}\\
                {0.6,}&{\frac{3}{4}N < i \le N.}
                \end{array}} \right.
			\end{equation}
			The setting of weights and thresholds are the same as the former one. 
			The result is shown in Fig. \ref{Fig:DNWLD_Pfa_Pd_PF_SNR2}, in which \ac{dwld} also controls the false alarm rate completely. 
            
		\subsection{Detection performance of \ac{dwld}}
		\label{subsec:detection_performance}
		
			\begin{figure}
				\centering
				\includegraphics[width=8cm]{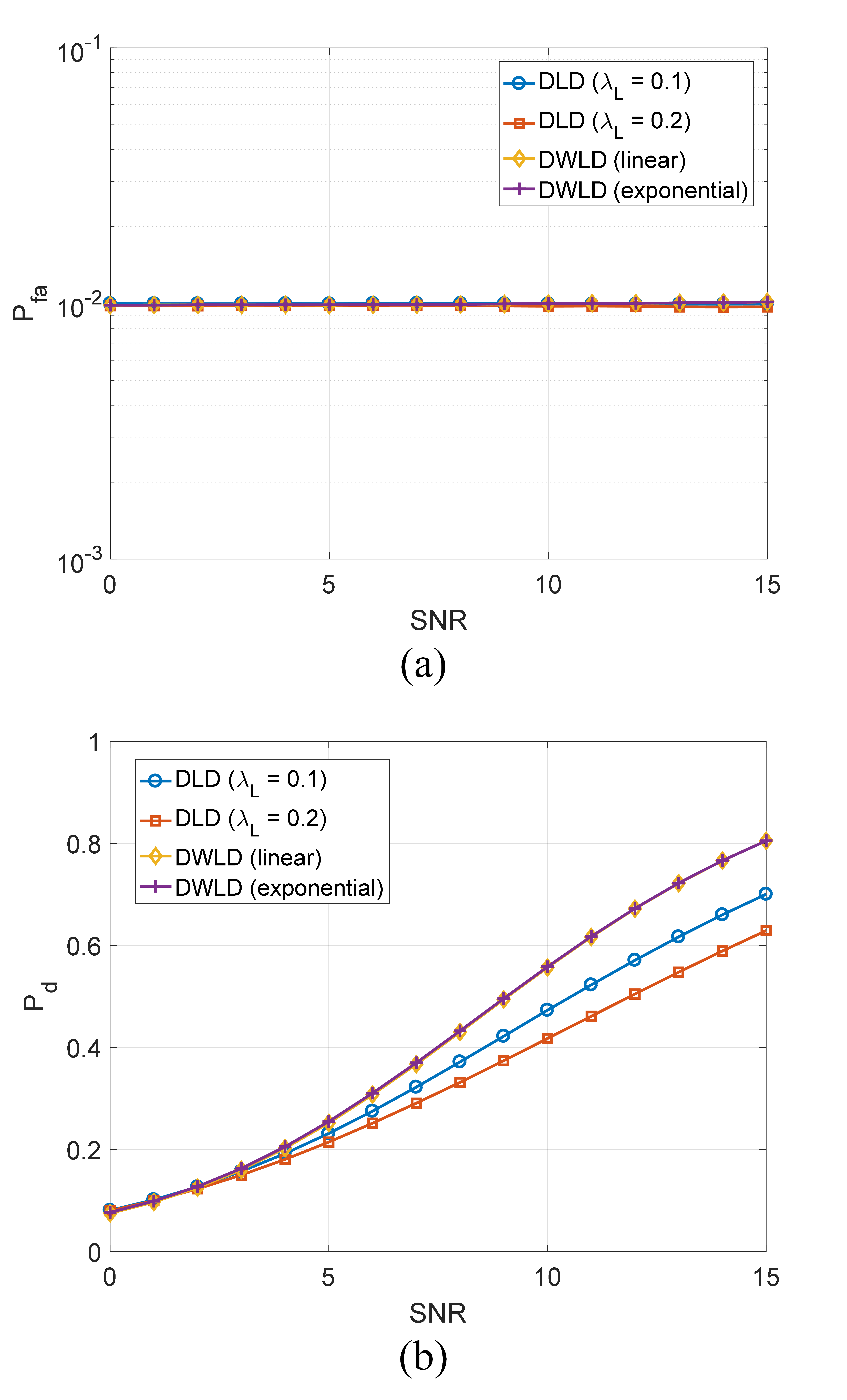}
				\caption{(a) Total false alarm rate and (b) total detection rate of different detectors vary with SNR for the first prior information setting \eqref{eq:pi1}. Here $\gamma = 0.5$.}
				\label{Fig:WL_Pfa_Pd_PF_SNR1}
			\end{figure}
			
			\begin{figure}
				\centering
				\includegraphics[width=8cm]{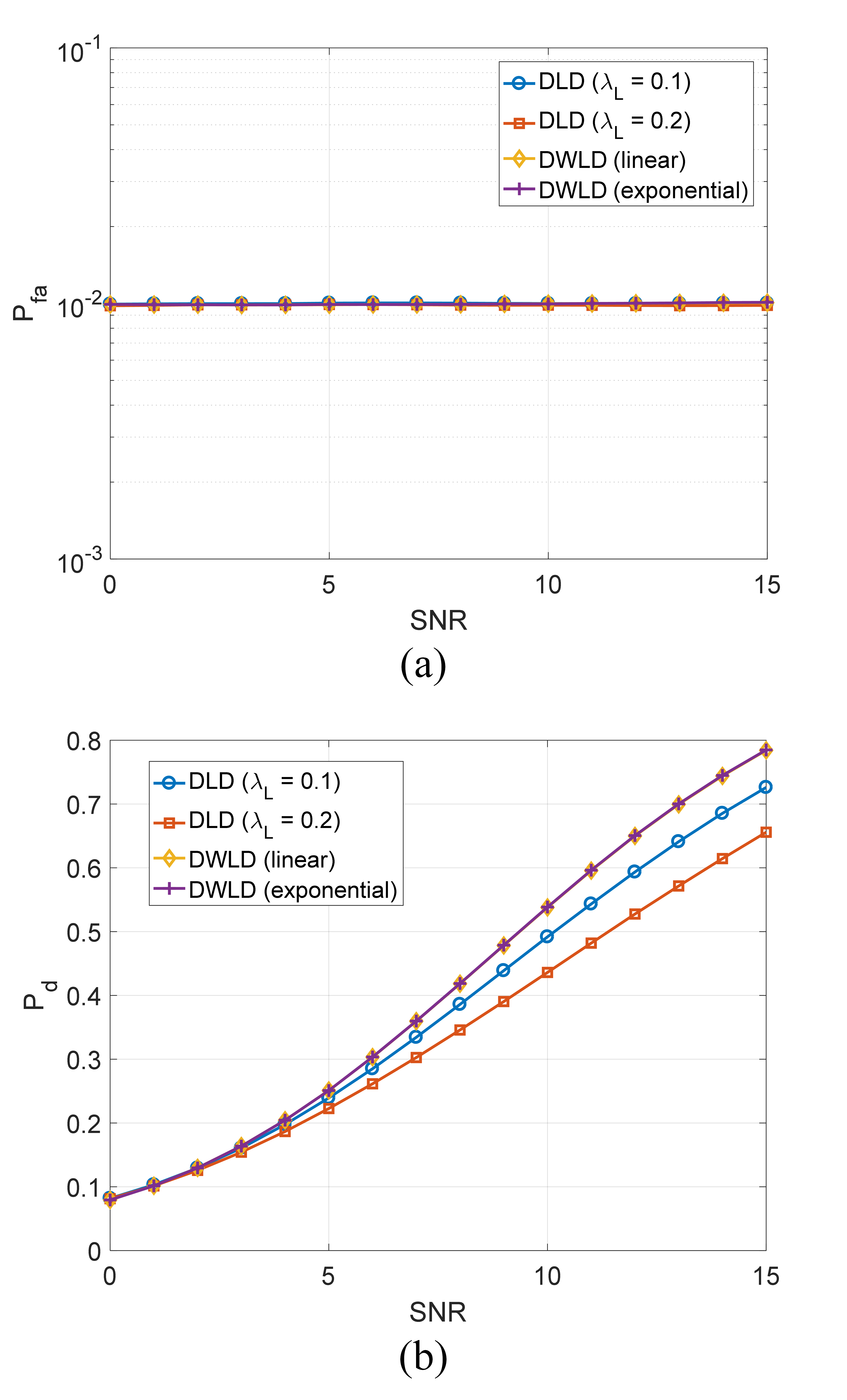}
				\caption{(a) Total false alarm rate and (b) total detection rate of different detectors vary with SNR for the second prior information setting \eqref{eq:pi2}. Here $\gamma = 0.5$.}
				\label{Fig:WL_Pfa_Pd_PF_SNR2}
			\end{figure}
			
			We then compare the detection performance of \ac{dwld} and \ac{dld}, and the parameter settings of $\bm A$, ${\bm x}_0$ and ${\bm \xi}$ are the same as Sec. \ref{subsec:comparison}. 
			The result of the first experiment is shown in Fig. \ref{Fig:WL_Pfa_Pd_PF_SNR1}, where the prior information $\bm p$ is setting as \eqref{eq:pi1}. 
			We compare the detection performance of \ac{dwld} with \ac{dld} for two regularization parameter settings. 
			The regularization parameters of \ac{dld} $\lambda_L$ are set to be $0.1$ and $0.2$.
			Meanwhile, the weight vectors $\bm \lambda$ were optimized to $\lambda_i = 0.1000 - 0.1002 p_i$ for the linear model and $\lambda_i = 0.0212 / p_i^{0.3470}$ for the exponential model for $\bm{p}$ of \eqref{eq:pi1}. 
			We demonstrate how the detection performance of these detectors varies with \ac{snr}. 
			The results suggest that both detectors can maintain the false alarm rate with different \ac{snr}, and the total detection rate of \ac{dwld} is higher than \ac{dld}. 
			
			The second one sets the prior information $\bm p$ as \eqref{eq:pi2}. 
			The regularization parameters of \ac{dld} $\lambda_L$ are set to be $0.1$ and $0.2$, and the optimal weight vectors $\bm \lambda$ were $\lambda_i = 0.0963 - 0.1038 p_i$ for the linear model and $\lambda_i = 0.0281 / p_i^{0.3594}$ for the exponential model. 
			The result is shown in Fig. \ref{Fig:WL_Pfa_Pd_PF_SNR2}, which indicates that \ac{dwld} also presents better detection performance than \ac{dld}. 
    
    \section{Conclusion}
    \label{sec:conclusion}
        
        The present paper introduces the design of the detector based on debiased weighted \ac{lasso} estimator for \ac{cs} radar systems. 
        The detection performance is analyzed theoretically and proved to be better than the naive one. 
        Aiming at improving the detection performance, we propose an approach to optimize the weights $\bm \lambda$. 
        By comparing \ac{dwld} with \ac{dld}, numerical results show that proper design of weights improves the detection performance.

	\bibliographystyle{IEEEtran}
	\bibliography{IEEEabrv,references}
					
\end{document}